\documentclass[graybox]{svmult}

\usepackage{type1cm}        
\usepackage{makeidx}         
\usepackage{graphicx}        
\usepackage{multicol}        
\usepackage[bottom]{footmisc}

\usepackage{newtxtext}       %
\usepackage[varvw]{newtxmath}       

\usepackage{titlesec}
\usepackage{amsmath}
\usepackage{amsfonts}
\usepackage{bm}
\usepackage{cancel}
\usepackage{changepage}
\usepackage{cite}
\usepackage{enumitem}
\usepackage{float}
\usepackage{helvet}
\usepackage{makecell}            
\usepackage{microtype}
\usepackage[numbers]{natbib} 
\usepackage{parskip}               
\usepackage[section]{placeins}
\usepackage{scrextend}           
\usepackage{subcaption}
\usepackage{url}
\usepackage{verbatim}            
\usepackage{xcolor}
\usepackage{xspace}

\usepackage[pagebackref=true,colorlinks=true]{hyperref}
\hypersetup{
     unicode=true,          
     pdfnewwindow=true,      
     bookmarksnumbered=true,
     colorlinks=true,       
     linkcolor=blue,          
     citecolor=blue,        
     filecolor=magenta,      
     urlcolor=cyan,           
     menucolor=blue,
     linktocpage=true
}

\def\TeV{\ifmmode {\mathrm{\ Te\kern -0.1em V}}\else
                   \textrm{Te\kern -0.1em V}\fi}%
\def\GeV{\ifmmode {\mathrm{\ Ge\kern -0.1em V}}\else
                   \textrm{Ge\kern -0.1em V}\fi}%

\makeindex             

\begin{document}

\title*{The Early History of the Quark-Gluon Plasma}

\author{Wit Busza and William A. Zajc}
\institute{Wit Busza  
   \at    Massachusetts Institute of Technology, Cambridge, MA 02139, USA;  
   \email{busza@mit.edu}
\and William A. Zajc
   \at    Columbia University, New York, NY 10027, USA; 
   \email{waz1@columbia.edu}}

\maketitle

\abstract{
We present the historical antecedents to the field of relativistic heavy ion physics, 
beginning with early attempts to model the strong interaction and ending with the endorsement of a 
relativistic heavy ion collider in the 1983 U.S. Long-Range Plan for Nuclear Science and the subsequent launch of heavy ion research programs at Brookhaven National Laboratory and CERN. Particular attention is paid to two major themes: 1) A program to study high density states of nuclear matter emerging from the 1974 Bear Mountain conference and 2) Efforts to understand the predictions of QCD for  matter at high densities and/or temperatures.
}

\section{Introduction}
\label{Sec:Intro}
Physics provides us with the tools and opportunities to study matter under extremes of pressure and density far removed from those found under ordinary terrestrial conditions. At the midpoint of the previous century, Enrico Fermi sketched the then-current understanding in a phase diagram labeled ``Matter in unusual conditions"~\cite{FermiThermoNotes}. The boundary of this diagram, corresponding to what Fermi described as an ``Electron proton gas'', was characterized by pressures $P \sim10^{28}$~atmospheres and temperatures $T \sim 10^{12}$~K. 

Presently, some seven decades following Fermi’s discussion, heavy ion collisions at the Relativistic Heavy Ion Collider (RHIC) at Brookhaven National Laboratory and the Large Hadron Collider (LHC) at CERN routinely create matter at temperatures larger than $10^{12}$~K ($\sim 100$~MeV) and pressures exceeding $10^{30}$ atmospheres ($0.6\  \mathrm{GeV}/\mathrm{cm}^3)$. 
Both experimental and theoretical  physicists  have developed exquisitely sensitive tools to study the particles and flow patterns emerging from each collision. 
These capabilities result from decades of advances in accelerator science, detector technologies, theoretical techniques and massively parallel computing facilities. 
A proper accounting of these efforts could well span multiple volumes. 
In contrast, this brief contribution provides a thematic overview 
of the early developments in experiment, phenomenology and fundamental theory 
that both motivated further research in the field and established the science case for doing so.

The field of study that today we call ``Relativistic Heavy Ion Collisions" or ``Quark Matter" has two independent origins. The first of these follows from the historical interest in high energy particle and/or nuclear collisions, tracing back to cosmic ray studies in the 1930's, together with the evolution of particle accelerators and eventually colliders. 
As described in Section~\ref{Sec:Exp}, this has been an evolutionary path, leading to our current capabilities to routinely perform sophisticated measurements on billions of events, each with thousands of final-state particles. 

The second development, which was much more revolutionary than evolutionary,
is related to both the motivation for studying nuclear matter under extreme conditions
as well as to shifting views of the very nature of that matter. In a brief span of roughly ten years, 
from 1973 to 1983, 
the focus shifted from exotic hadronic states to an entirely new form of matter, 
the quark-gluon plasma. 
In Section~\ref{Sec:Th} we identify some of the seminal events and contributions~\cite{Methods} that led to this paradigm shift. 

This review is organized around those two themes, the first being the long, essentially continuous evolution of experimental observations of particle production via the strong interaction discussed in Section~\ref{Sec:Exp}, along with the relevant pre-QCD phenomenology, from the 1930's to the early 1980's. This is in contrast to the relatively short but hugely impactful period of intense theoretical progress from 1973 to 1983 detailed in Section~\ref{Sec:Th}. Section~\ref{Sec:1983} describes how these parallel developments converged in 1983 to firmly establish the basis for the field of relativistic heavy ion physics. 

\section{Experimental Study of Multiparticle Production}
\label{Sec:Exp}
\subsection{Historical Antecedents}
\label{Sec:History}
One of the origins of the field can be traced back to  the 1930's;
neutrons  and nuclei had been discovered.
A long-standing controversy on the primordial\footnote{Here ``primordial" refers to the composition prior to showering in the atmosphere} composition 
of cosmic rays had been resolved: they are dominated not by high-energy photons, 
but rather by
ultra-relativistic protons and nuclei~\cite{WikiCosmic}. Effectively nature has provided a low luminosity but high energy proton and nucleus accelerator for studies of how such particles interact with matter.

As strange as it may seem to the modern reader, due to the prevalence of QED-based studies,
initially there was considerable skepticism that multiple particles could be produced
in a single collision. 
These reservations were laid to rest by observations clearly establishing
the production of greater than 3 particles in a single collision, including even  a spectacular 8-particle ``star"  in an emulsion placed on  a mountain top~\cite{Blau1937}.
These results provided direct evidence that short-range strong forces exist
and that they were indeed strong, demonstrating that 
electromagnetic fields are not the only fields that exist and that radiation is not the only mechanism of particle production.

Surrounding strongly interacting particles there must exist strong fields with their own corresponding carriers.
Questions arise regarding what happens when strong fields collide? Heisenberg~\cite{Heis39} asks: is it a slow process or a fast one? If it is a slow process, the time of production of particles must be much longer than the collision time.
In that case what is the state that exists during the ``slow" production process?

In the 1960's and 1970's there is interest in the formation time of a produced particle~\cite{Gottfried:1974yp,Bialas:1974qs} and how the production process evolves in spacetime, including how the newly formed particles interact.
Various models are invented to explain multiparticle production data; e.g., the one and two-fireball models~\cite{Cocconi:1958zz,Akimov:1969buq,DeTar:1970jsi}

Early in the 1960's Glauber develops the  ``Glauber Model", a quantum mechanical analysis of the relation between elastic scattering off nuclei to that off the individual nucleons~\cite{Glauber:1970jm}. It is first applied in the study of the elastic scattering of p+p and p+A collisions, then to the coherent production of vector mesons in $\gamma$+A collisions (ideas redeployed in today’s studies of ultra-peripheral collisions).
In the late 1960's / early 1970's the Glauber Model is developed by Bialas, Bleszynski,
and Czyz~\cite{Bialas:1976ed,Bialas:1977pd} to form the basis of the Glauber Method, currently used by the community for determining the impact parameter in p+A and A+A inelastic collisions~\cite{Miller:2007ri}. 

In the early 1970's the question is raised ``Do p+p collisions teach us about p+A collisions or vice versa?", and Gottfried introduces his  influential Energy Flux Cascade Model~\cite{Gottfried:1974yp}. It is uncanny how the energy flux properties elucidated by Gottfried more than 50 years ago are similar to today's quark-gluon plasma analyses.

Most of the  experimental data in the 1960's was still from studies of cosmic ray interactions with nuclear emulsions. 
However, there is  a big problem with the interpretation of such data; neither the energy nor particle type of the incident cosmic ray,  nor the nature of the target nucleus are well determined.
In addition, only the angles with respect to the initial cosmic ray direction are determined; particle identification is not available for the produced tracks. It was in this context that pseudorapidity~\cite{pseudo}, now ubiquitous in the field, was introduced as a variable of necessity. Fortuitously, it was also found to have a useful {\em dynamical} context as an appropriate variable to describe the longitudinal phase space in these relativistic collisions. 

A major advance in these studies was made by Jones and collaborators in a cosmic ray experiment at Echo Lake which solved one of the problems, resolving for each collision
the nature of the target~\cite{Vishwanath:1975dy}. Studies of particle production with known beams and targets had to wait for the first truly high energy proton beams, first at 70 GeV from the Serpukhov U-70~\cite{bougorski_2024_mykfw-xbe41}, followed by  the 200 GeV beams from the U.S. National Accelerator Laboratory (Fermilab) in 1972~\cite{NAL} and 400 GeV proton beams from the CERN SPS in 1976~\cite{CERNSPS}.

In the 1970's the experiments at Fermilab are game changers. For the first time analyses of significant numbers of events, with known beams and targets, allow quantitative comparison of models with data.
At a qualitative level it is shown that in p+A collisions there is no cascading of produced particles inside the target nucleus and there is some kind of transparency for the very forward-going particles~\cite{Whitmore:1976ip,Chaney:1977ja,Elias:1979cp}; 
the physical picture being more akin to the modest multiplicity growth from colliding liquid drops breaking up into further liquid drops rather than the 
cascade expected from successive generations of shattering glass spheres. At a quantitative level, participant and extended longitudinal scaling~\cite{Dao:1974sp} are discovered.
As is the case today, much effort was spent in trying to understand in detail what are the various mechanisms of multiparticle production, from before the collision, through the impact of the two incident particles, to the  evolution and decay of the produced intermediate state, to the final production of the asymptotically produced particles.

In 1971, the first hadron collider, the CERN Intersecting Storage Rings (ISR)~\cite{ISR} began operations with 
p+p collisions at $\sqrt{s} =53$~GeV. 
From the standpoint of collider technology, the ISR had a major impact with many firsts,
including the first demonstrations of RF beam stacking and stochastic cooling. 
From the standpoint of physics impact, 
the ISR presents a fascinating object lesson, in that the theoretical prejudices 
at the time it was built were encoded in the concrete and steel of the initial suite of detectors,
all of which concentrated on the forward/backward regions. As a result, 
the discoveries of large transverse momentum production~\cite{Carey:1974fk,Cronin:1974zm} 
and the $\mathrm{J}/\psi$~\cite{SLAC-SP-017:1974ind,E598:1974sol}
were made elsewhere. 
For further information on this perspective we refer the reader to Darriulat's charming and nuanced account
of the ISR history~\cite{Darriulat:2012yf}.
Eventually detectors were built specifically for the study of mid-rapidity phenomena at the ISR, 
pioneering technologies such as full hadronic calorimetry that are in routine use today~\cite{Giacomelli:1979nu,Jacob:1984dy,Fabjan:2004qf}. 
Another first at the ISR was the initial studies of nuclear collisions
at a collider, albeit  limited to p+$\alpha$, d+d and $\alpha+\alpha$ collisions~\cite{Jacob:1981zb}.
Nonetheless, these early studies included analyses on strangeness production~\cite{AXIALFIELDSPECTROMETER:1985eyj}
and wounded nucleon convolution methods~\cite{Callen:1986kb,ANGELIS1984140}
for transverse energy production.

At the end of this discussion of one of the two threads of research that led to the current field of relativistic heavy ion collisions it is worth mentioning a few more facts. In the late 1970's /early 1980's there was interest in the degree to which nuclei were transparent to the produced forward-going high energy particles. A by-product was the measurement of the so-called nuclear stopping power and the first determination of the energy density that will be produced in the RHIC collider which was in the planning stage at the time\cite{Busza:1983rj}. The results indicated that the energy density that will be produced at RHIC should be interesting. In short, RHIC will not be a bust!

Two other results were also by-products when nuclei were used as targets, in the place of protons (to increase the event rate of collisions). One was the discovery of the Cronin effect~\cite{Cronin:1974zm} and the other of the EMC effect~\cite{EuropeanMuon:1983wih}. The understanding of these effects remains important 
in current analyses of pA and AA collisions at RHIC and the LHC.

Finally, while to this point we have stressed the insights obtained from proton-nucleus collisions, we should also emphasize here the crucial role played by experimental programs with nuclear beams, initially at the Bevalac~\cite{Grunder1973heavy,Bevatron} which from 1974 to 1993 provided beams of heavy nuclei with 1-2~GeV/nucleon. Those early studies in the late 1970's and 1980's developed many of the fundamental techniques in the field, ranging from single-particle spectra to early flow phenomena to HBT methods and dilepton measurements. Subsequent work, at higher beam energies but still in fixed target mode, was performed at the Brookhaven AGS and the CERN SPS, with both programs beginning in 1986, both serving as critical preludes to the collider programs to come. 

\subsection{Relevant pre-QCD Strong Interaction  Phenomenology}
\label{Sec:Phenom}

By the late 1940's it was becoming clear  that the perturbative expansions
being developed for QED would fail for the strong interaction, precisely because it was strong and the higher order terms were at least as important as the first-order terms. In response to this, in 1950 Fermi developed a statistical model\cite{Fermi:1950frz}
he described as ``extreme, although in the opposite direction, as  the perturbation theory approach". 
Accordingly, he assumed that all of the energy in a nucleon-nucleon collision is deposited and statistically equilibrated in a volume $\Omega$, which is the only adjustable parameter in the theory. The resulting number of particles and their energy distributions are then those of $N$-body phase space, which Fermi evaluates for various energies of interest at that time.  

While it was clear from Fermi's text that this model was intended as an upper limit for the expected particle production, the consistency of its assumptions was soon questioned. Pomeranchuk~\cite{Pomeranchuk:1951ey}  noted that Fermi failed to account for the interactions between the produced particles and the subsequent expansion of the reaction volume. Soon thereafter, Landau\cite{Landau:1953gs} and then Belinkij and Landau~\cite{Landau:1955tlt} provided a complete dynamical analysis of Pomeranchuk's suggestion in the framework of relativistic hydrodynamics. 

The subsequent history of applying hydrodynamics to describe multiparticle production in hadron-hadron collisions is very curious. The great majority of researchers chose not to deploy this approach, perhaps reflecting the sentiment expressed by I.I.~Rabi at the 1949 Chicago Meeting of the American Physical Society, where he stated ``If Fermi is right in saying that he can calculate what will happen at very high energies by purely statistical methods, 
then we will have nothing new to learn in this field."\cite{RabiOnFermi}.  
Acceptance of the hydrodynamic approach required, in effect, abandoning
the prevailing {\em zeigeist} that particles, described as excitations of a quantum field, were the uniquely interesting field excitations to be studied.
Despite the hydrodynamic model's success in describing the broad features (rapidity distributions, transverse momentum distributions, multiplicities) of inclusive hadron production, 
only a small dedicated band of practitioners~(see \cite{Carruthers:1973ws} and references therein)
actively pursued Landau's approach. The situation circa 1981 was neatly summarized by Carruthers' characterization of the prevailing attitudes as ``the strong interaction is too hard to solve, but the hydrodynamic theory is too hard to accept"~\cite{CarruthersQuote1981}. 

Another, initially independent, line of inquiry led to phenomenological models of the energy deposition process when a hadron undergoes multiple collisions. The methods here were very much driven by the experimental data discussed in Section~\ref{Sec:Exp}, in particular for p+A collisions, which proved to be an ideal laboratory for understanding the {\em lack} of cascading in the proton-going direction. Work by Gottfried~\cite{Gottfried:1974yp}  and by
Bialas and Czyz~\cite{Bialas:1974qs} indicated that this could 
be understood in terms of the formation time for secondary particles. 
The latter authors also applied the Glauber model~\cite{Glauber:1955qq} 
based on diffraction theory and its extension to inelastic collisions~\cite{Glauber:1970jm}
to determine the number of ``wounded nucleons"~\cite{Bialas:1977pd}; this technique with its extension to nucleus-nucleus collisions~\cite{Bialas:1976ed} would become hugely influential in the subsequent development of space-time models of heavy ion collisions~\cite{Miller:2007ri}. 

Indeed, the very fact that both the hydrodynamic models and those describing the energy flux associated with multiple collisions were explicitly formulated in
space-time, following both the spatial and temporal development of particle production, represented a challenge to the then-extant models of hadron production. 
Workers in the field today may not appreciate the extent to which this ran counter to the prevailing approaches at the time, whether based in quantum field theory, the S-matrix program~\cite{Chew:1962mpd,Eden:1971jm} or more heuristic prescriptions such as the parton or multiperipheral models~\cite{PhysRev.176.2112,Silverman:1971rq,Koplik:1975ni,Steinhoff:1974zw}. 
The interested reader will find a prescient discussion by Bjorken on this topic in his 1975 SLAC Lectures~\cite{Bjorken:1976mk}. The importance of moving the discussion from momentum space to phase space\footnote{Note that the discussion moved {\em back} from (only) momentum space to phase space; the early work by Fermi and Landau made clear reference to the coordinate space features we now term peripheral and central collisions.}
cannot be overemphasized. 
It forms the very basis for modern, defining aspects of relativistic heavy ion physics such as hadronic energy and/or entropy densities and their time evolution,
transverse shapes and sizes, flow fields, jet quenching, etc. 
These in turn evolved in response to critical developments in experimental proton-nucleus studies~\cite{Busza:1975te}, which demanded the understanding of spatial-temporal concepts such as stopping and cascading. 

To conclude this section, we would be remiss not to mention the statistical model of Hagedorn and co-workers. 
It is far removed from a theory describing the space-time of multi-particle production, 
rather it aims to understand
a striking result from the end product of that production - the observed exponential increase of  the density of states $\rho(m)$ with mass $m$. Motivated by this, Hagedorn developed an elegant ``statistical bootstrap" theory~\cite{Hagedorn:1965st} based on the assumptions that the strong interaction is as strong as is allowed by unitarity, and that this results in an infinity of resonances  (``fireballs") with the observed exponential mass density that can be described by statistical mechanics, with the important bootstrap condition that ``fireballs consist of fireballs". This led to a successful description of transverse momentum distributions $ \frac{dn}{p_T dp_T} \sim \sqrt{m_T} \ e^{-m_T / T_0}$ with 
$m_T \equiv \sqrt{p_T^2 + m^2}$ and $T_0 =158 \pm 3$~MeV. Crucially, self-consistency of the theory necessitated that $T_0$ should be understood as the highest temperature allowed in nature, 
a startling conclusion that led to conundrums for theories of the early universe~\cite{Huang:1970iq}, a problem that persisted until the advent of QCD described in the next section. 

\section{Theoretical Developments}
\label{Sec:Th}
In some very real sense, one of the main motivations  
to investigate high energy collisions of nuclei arose out of ``the study of nothing'',
in particular the 1974 paper by Lee and Wick\cite{Lee:1974ma} exploring the implications
of altering the vacuum expectations of a scalar field over a length scale 
``substantially greater than the usual microscopic dimension in particle physics''. 
A direct line may be traced from this work, published in April of 1974 to the November 1974 Bear Mountain workshop, where Lee's talk  {\em A Possible Form of Matter at High Density} (available as \cite{Lee:1974nj})\footnote{This obscure method of publication 
has resulted in only one citation, but there can be little doubt as to the impact of Lee's proposal on the 
high energy nuclear physics community.}
called for the study of bulk hadronic phenomena in events
where the collision energy is deliberately distributed over a relatively large volume. 

There is no doubt that the Bear Mountain conference was a landmark event, 
bringing together formerly disparate elements of the nuclear physics, high energy physics, astrophysics and accelerator science communities to consider the science that
would be enabled by high energy collisions of heavy nuclei. 
However, it must be viewed in historical context. 
The discussion at the conference focused on the theoretical frontiers of nuclear physics at that time, 
such as pion condensation, nuclear shock waves and the stability of strange nuclei. 
The words ``quarks" and ``gluons'' are nowhere present in the proceedings. 
Nor should we be surprised - the theory of quantum chromodynamics (QCD)
was in the process of being invented at precisely the same time as the workshop. 
For perspective, it is worth noting that neither ``quark'' nor ``gluon'' appears in Politzer's 1973 paper on reliable perturbative results from the strong interaction\cite{Politzer:1973fx}. Similarly, the 1973 paper by  Gross and Wilczek
on asymptotic freedom in gauge theories\cite{Gross:1973id} refers only to quarks in closing, again with no reference to gluons. 

These early QCD papers, followed by more expansive work 
by the same authors in 1974\cite{Politzer:1974fr,Gross:1974cs}, 
were the key steps in identifying quarks and gluons as {\em the} dynamical degrees of freedom for the strong interaction, rather than simply as the group theoretical underpinnings 
of the structure of hadron multiplets in the quark model as discovered by Gell-Mann\cite{Gell-Mann:1964ewy} and Zweig\cite{Zweig:1964ruk,Zweig:1964jf}. 
Prior to the discovery of QCD, some authors such as Itoh\cite{Itoh:1970uw} considered the implications of promoting quarks to dynamical objects for stellar structure,
but this had little impact due to the schematic nature of the model used. 

The 1975 work of Collins and Perry was more impactful precisely because it was fully informed by then current knowledge of the gauge structure of QCD. They demonstrated that a perturbative treatment is possible at baryon densities not much larger than normal nuclear density $\rho_0$, with asymptotic freedom becoming increasingly valid as  the density increases. They noted that in both the high baryon density regime and that of high temperature (independent of baryon density) many-body effects produce an effective mass for the gluons, thereby ameliorating the infrared problems expected in Yang-Mills theory of massless gauge bosons. Similar calculations were performed in 1977 by Chapline and Nauenberg\cite{Chapline:1976gy}, who found that, while qualitatively reproducing results from the phenomenological MIT bag model\cite{Chodos:1974je}, 
a phase transition to a deconfined state would require densities as high as 10-20$\ \rho_0$, thereby casting doubt on relevance to the inner core of neutron stars. 
This result more firmly established a similar conclusion obtained by Baym and Chin\cite{Baym:1976yu} in 1976 working directly in the MIT bag model. 
For our purposes, it is notable that while the authors of these papers cite possible applications to neutron stars, the early universe and black hole explosions (Hawking radiation), no mention is made of experimental study via collisions of heavy ions. 

Arguably more important to the eventual focus of heavy ion collisions was the understanding of thermal properties of gauge theories. 
Even prior to the advent of QCD as the theory of strong interactions, 
Dolan and Jackiw studied the thermal properties of gauge theories\cite{Dolan:1973qd}, demonstrating that restoration of broken symmetries
could appear above a critical temperature. 
By 1976, that  work was extended and applied to the explicit case of QCD by 
Kislinger and Morley\cite{Kislinger:1975uy,Kislinger:1975ab}, who showed that
plasmons with mass $\sim gT$ must exist in non-Abelian theories, 
thereby eliminating the possibility of long-range forces at finite 
temperature even in asymptotically free theories, an observation 
of great importance to both the then-developing theory of the first 
microseconds of the early universe and (eventually) to quark-gluon plasmas studied in heavy ion collisions. 

Curiously, the statistical bootstrap model
discussed in Section~\ref{Sec:Phenom}
provided another suggestion that 
hadronic matter undergoes a phase transition at high temperature.
Already in 1975 Cabbibo and Parisi~\cite{Cabibbo:1975ig} argued that the 
limiting temperature in the Hagedorn model was in reality the signature 
of a second-order phase transition to a state where
``quarks can move throughout space,” that is, were deconfined. 
Later work by Hagedorn and Rafelski~\cite{Hagedorn:1980kb} argued 
that this would in fact be a first-order phase transition. 
While today, we know that at zero net baryon density this is in fact
a (crossover) phase transformation rather than a finite-order
phase transition, it is nonetheless quite remarkable 
that the self-consistent theory of resonances exhibiting an exponential density of states predicts not only its own demise but also rather accurately
the temperature $\sim 160$~MeV at which this should occur. 

By 1978, developing insights into experimental possibilities begin to appear. For example, in their work on quark star phenomenology, Freedman and McLerran~\cite{Freedman:1977gz} mention in their first paragraph the possibility of a nuclear phase transition not only in the core of a pulsar, but also in heavy ion collisions\footnote{These authors, in their Reference 9, in turn credit R.~Jaffe and A.~Kerman for this suggestion. We in turn note that Kerman participated in the Bear Mountain conference and is likely to have brought those discussions to the attention of Freedman and McLerran.}. 
Yet substantial reservations remained, as evidenced in David Gross's talk at LBL's {\em 1st Nuclear Workshop on Ultra-Relativistic Nuclear Collisions}~\cite{QMconf} in 1979, where he cites concerns ``emphasized by many of the
speakers at this meeting" as to whether high density states will be created at all in relativistic heavy ion collisions, and if they are formed, whether there will be adequate time to establish thermal equilibrium\cite{Gross:1979zq}. 
These reservations
are completely understandable. After all, the key property of QCD was asymptotic freedom, which provided an all too convenient explanation of the leading particle effect discussed in Section~\ref{Sec:Exp} - higher energies (it was thought) would lead to nuclear transparency due to the reduced interactions between the quarks at large momentum transfer. During this same period the role of gluons in defining the vacuum structure of QCD was intensively studied\cite{Belavin:1975fg,Callan:1976je,Callan:1977gz,Gross:1980br,Shuryak:1981ff},
indicating that the QCD incarnation of T.D. Lee's oft-repeated dictum 
to explore the physical vacuum could be found in the topological vacuum fluctuations and instantons of the theory. 

These difficulties seem to indicate that the central role (pun intended)
of the gluons in the nucleon had not been fully assimilated by the community.
As early as 1975, van Hove and Pokorski\cite{VanHove:1974wa} argued that particle production in the 
mid-rapidity (aka ``central") )region of hadron-hadron collisions was dominated by gluon-gluon collisions. 
Building on this observation, Shuryak\cite{Shuryak:1976ny} in 1976, made a clear distinction between the leading particles, interpreted as the fragmentation of the valence quarks, and the high energy density region of ``neutral glue" formed by the gluons. 
It was also in this early period that Shuryak, working essentially in isolation (literally in Siberia) elucidated the properties not only of baryonic rich QCD matter but also in the high temperature but baryon free central region\cite{Shuryak:1977ut}, clearly demonstrating the screening of the Coulomb field in such a medium. 
Subsequently, 
considerable effort was invested by various authors\cite{Linde:1978px,Kapusta:1979fh,Kalashnikov:1979cy,Kalashnikov:1979dp,Linde:1980ts,Shuryak:1980tp} to understand the thermal properties of QCD, in particular, the higher-order corrections and 
the nature of the phase transition.

Intriguingly, during this period the theory of interacting gluons was being developed in a parallel path initiated by Wilson's 1974 prescription for quantizing a gauge field theory on a discrete lattice in Euclidean space-time \cite{Wilson:1974sk}, with a particular emphasis on  understanding the confinement of quarks. 
Wilson's  work, which appears to have been motivated in part to provide a
more rigorous basis to support the qualitative arguments from Kogut and Susskind\cite{Kogut:1974sn}
was followed shortly by another Kogut and Susskind paper\cite{Kogut:1974ag} providing a Hamiltonian basis for Wilson's lattice methods.
An independent argument by Polyakov\cite{Polyakov:1978vu} in 1978
used lattice-based methods to motivate a quasi-analytic argument
that QCD has a phase transition to a deconfined state where gluons
acquire a non-zero mass. 
By 1979, Susskind\cite{Susskind:1979up} had demonstrated that the confining properties of QCD at zero temperature were lost at sufficiently high temperature, resulting in a deconfined phase characterized by Debye screening due to a plasma of gluons. 

An absolutely key step in these lattice-based investigations of the strongly-coupled limit of QCD was Creutz's 1980 demonstration\cite{Creutz:1980zw}  that numerical computations using the Metropolis algorithm exhibited both confinement and asymptotic freedom in appropriate limits. While the limits of computational resources restricted these studies to $SU(2)$ models, Creutz's distribution of card decks\footnote{A deck was a set of punched cards for entering source code (and in some cases data), typically with one line of FORTRAN per card} to interested parties enabled many researchers to 
deploy these methods; an early laudable instance of the open source movement in scientific computing.
Within a year 
McLerran and Svetitsky\cite{McLerran:1980pk,McLerran:1981pb}, 
Kuti, Polónyi and Szlachányi\cite{Kuti:1980gh},
and
Engels, Karsch, Montvay and Satz\cite{Engels:1980ty,Engels:1981qx}
used these techniques to provide illuminating results on the nature of the phase transformation to deconfined gluons along with its properties such as the transition temperature, specific heat, the $q$-$\bar{q}$ potential and the string tension.

It was also in 1980 that Shuryak suggested that
the deconfined state of QCD at high temperatures and densities be referred to as a ``quark-gluon plasma"~\cite{Shuryak:1980tp}. 
Previous researchers had proposed 
quarkium, quarkionic matter, quark soup, hadronic plasma, quark-antiquark plasma, etc., but none of these are as evocative of the true nature of this state as the phrase quark-gluon plasma (QGP)~\cite{sQGP}. 
Since it is indeed the non-Abelian nature of the gluon interaction that imbues QCD with its unique properties of a negative $\beta$-function, leading in turn to asymptotic freedom and the possibility of a deconfined state, it seems only appropriate that the name of this new form of matter should reflect this key constituent. 

Finally, theoretical interest in high density and/or high temperature 
hadronic matter may have waned were it not possible to explore 
these conjectures experimentally. 
As described in Sections~\ref{Sec:History} and \ref{Sec:Phenom}, 
progress was driven by the detailed studies of proton-nucleus collisions.
Substantial effort was required to reconcile
the confounding observations of limiting fragmentation and longitudinal scaling with energy deposition associated with the growth of the gluon-dominated central rapidity region~\cite{Busza:1975te}.
The increased understanding of nuclear stopping also led to a shift of
attention from the forward fragmentation regions with their high net baryon density\cite{Anishetty:1980zp} to the nearly baryon-free gluon-dominated central regions made possible by higher collision energies; 
we cite here but one example\cite{McLerran:1981ue} of this general trend.
This understanding led in turn to predictions for new experimental 
signatures from gluon-gluon collisions, in particular,
the very influential 1982 prediction by M\"{u}ller and Rafelski~\cite{Rafelski:1982pu} of greatly enhanced strangeness production via this process.
Further experimental developments\cite{Willis:1981xm} towards the end of this period played a significant role in guiding theoretical interest, in particular, the possibility of pursuing  truly high energy ($\sqrt{s_{NN}} > 30$~GeV) light-ion collisions at the CERN Intersecting Storage Ring (ISR) and the prospects for a major advance in the energy of p+p collisions at Isabelle, a high-energy physics collider under construction at Brookhaven National Laboratory.

\section{1983 - {\em Fundamentum Anni} (Foundational Year)}
\label{Sec:1983}
For the field of relativistic heavy ion physics, 1983 was a remarkable year, marked by confluence, convergence and consequence. 
More specifically, that year saw confluence in a particular theoretical approach to applying hydrodynamics, 
convergence of experimental and theoretical efforts in understanding the expected reaction dynamics, and far-reaching consequences of
funding decisions made by the U.S. government. 

Confluence was present in Bjorken's 1983 seminal publication 
{\em Highly Relativistic Nucleus-Nucleus Collisions: The Central Rapidity Region}~\cite{Bjorken:1982qr}, which combined Landau's hydrodynamic theory with the novel {\em ansatz} of boost invariance for the initial conditions,  motivated in large part~\cite{BjNote} by 
the data and models 
ranging from Feynman's parton model~\cite{Feynman:1969ej} 
to the phenomenology of stopping in proton-nucleus collisions~\cite{Busza:1975te,Busza:1983rj}.
By doing so, and by treating only the longitudinal motion, Bjorken was able to produce an exact solution to the hydrodynamic equations of motion~\cite{preBj}, 
which in turn led to a pocket-formula for estimating the energy density 
obtainable in nuclear collisions. Application of this result to future facilities suggested that a center-of-mass energy per nucleon pair
$\sqrt{s_{NN}}$ of $\sim 100$~GeV would be sufficient to form quark-gluon plasma. 

The convergence that we refer to is that of theoretical and experimental interests of the trends discussed in the preceding material - the decades-long series of experimental investigations first performed with cosmic rays then at accelerators discussed in Section~\ref{Sec:Exp}, in conjunction with a decade of intense theoretical development from 1973-1983 described in Section~\ref{Sec:Th} aligned to form a compelling case
for a new facility. 
The arguments, based on the realization that gluons were an essential component of the quark-gluon plasma, and that the most likely place to study high temperature deconfined nuclear matter with its full complement of gluonic degrees of freedom was in the central region (mid-rapidity) of high energy nuclear collisions, established the case for a high energy nuclear collider. 
While it was possible that such conditions could be produced in proton-proton collisions (as noted explicitly in Bjorken's hydrodynamic paper), there was general agreement that the possibilities of successfully producing statistically equilibrated matter would be maximized in central collisions of heavy nuclei, fulfilling T.D. Lee's dictum from the 1974 Bear Mountain conference~\cite{BearMountain}:
\begin{quote}
    In order to study the question of the ``vacuum", or the possibility of abnormal states, we must turn to a different direction; we should investigate some ``bulk" phenomena by distributing high energy or high nucleon density over a large volume. {\em The fact that such directions have never been explored should,} by itself, {\em serve as an incentive for doing such experiments.}
\end{quote}

Finally, it is through no small measure of serendipity that the consequences of strategic decisions made separately in the U.S. high energy physics and subsequently in the nuclear physics community made possible the construction of an accelerator complex and set of experiments dedicated to precisely the program envisioned by Lee and others. In 1974, the U.S. High Energy Physics Advisory Panel (HEPAP) had recommended that a proton-proton collider, ISABELLE~\cite{ISABELLE}, with center-of-mass energy 400 GeV, be constructed at Brookhaven National Laboratory (BNL). The latter rechristening of the project as CBA (Colliding Beam Accelerator) could not overcome the initial delays in developing reliable superconducting magnets. This realization, together with the enormous successes of CERN's proton-antiproton collider, rendered the major goals of CBA moot, leading to the cancellation 
of the project by the U.S. Department of Energy in July, 1983~\cite{CBA}.

The termination of CBA could have struck a fateful blow
to BNL's aspiration to construct a collider. However, an enlightened BNL management team, led by BNL Director Nick Samios and in consultation with T.D.~Lee (Columbia) and Jack Sandweiss (Yale), fortunately had made contingency plans for CBA to be re-purposed as the Relativistic Heavy Ion Collider (RHIC), a facility capable of 
colliding species as heavy as Au ions at $\sqrt{s_{NN}} = 200$~GeV.  
Providentially, the Long-Range Planning Committee for Nuclear Science was meeting that summer in upstate New York to establish priorities in nuclear physics for the next decade. A fascinating description of the 
subsequent ``shuttle diplomacy" that ensued, with BNL's Tom Ludlam acting as envoy between lab management and the planning committee,  may be found in Ref.~\cite{CreaseRHICBirth}, an effort that led to one of the three major recommendations of the 1983 Long-Range Plan:
\begin{quote}
    We identify a relativistic heavy ion collider as the
highest priority for the next major facility to be
constructed, with the potential of addressing a new
scientific frontier of fundamental importance.
\end{quote}
It is critical to emphasize here that this was much more than
recognition of an unanticipated opportunity. Rather, the recommendation to invest in a relativistic heavy ion collider represented the culmination of decades of experimental and theoretical investigations described in Sections~\ref{Sec:Exp} and \ref{Sec:Th}.
As noted in the preface to the Quark Matter 1983 conference~\cite{Ludlam:1984flw}
(coincidentally held at BNL in September of 1983), 
key among these was the demonstration, based on 
proton-nucleus data and phenomenological analysis~\cite{Busza:1975te,Busza:1983rj},
that the center-of-mass energies accessible at RHIC would provide a clear separation between the baryon-rich fragmentation regions and the central-rapidity region characterized by high energy density with low net baryon density, ideal conditions for analyzing the fascinating properties of the quark-gluon plasma. 

RHIC was approved for construction in 1991, and began operations in  2000. The discoveries that flowed from this facility, the most flexible collider ever built, are described throughout this volume. Those striking insights relied in an essential way not just on a new collider providing the requisite center-of-mass energies, but also on a suite of four complementary experiments capable of performing precision measurements in regimes of unprecedented particle densities~\cite{RHIC4Exp}. The success of RHIC stems from this confluence of a dedicated collider in consort with a carefully optimized experimental program.

\section{Summary}

The endorsement of RHIC in the 1983 Long Range Plan for Nuclear Science marked ``the end of the beginning" for the field of relativistic heavy ion physics. As emphasized here, the beginning was a prolonged period stretching over four decades~\footnote{We have provided an annotated history of these decades in the \hyperlink{Appendix}{Appendix}.}, 
as researchers struggled with the very features that made the strong interaction strong: multiple particle production, the failure of perturbative treatments and the importance of understanding the space-time development of energy deposition. By 1983, an enormous intellectual effort, stimulated by the formulation of QCD in 1973-4 and essential experimental  studies in proton-nucleus collisions, established the intellectual validity and open questions in the field. Addressing those questions both theoretically and experimentally, define the next chapters in this history, and are detailed elsewhere in this volume. 
While, after a quarter-century of spectacular success, the RHIC program will transition to new investigations at the Electron-Ion Collider, the LHC heavy ion program at CERN will continue for the foreseeable future. It is clear that through these experimental efforts at BNL and CERN, and the attendant theoretical descriptions of the phenomena, the field has succeeded spectacularly in the study of ”matter under unusual conditions” envisioned by Fermi more than seven decades ago.


\setcounter{secnumdepth}{0}
\section{Acknowledgments}
\label{Sec:Ack}

We wish to acknowledge essential conversations with and suggestions from\\
T.~DeGrand, M.~Gyulassy, J.~Kapusta, L.~McLerran, B.~Mueller, J.~Nagle, K.~Rajagopal and E.~Shuryak. We also wish to acknowledge Tapan Nayak's careful and thorough editing, which has greatly improved our manuscript. 
This work was supported by the U.S. Department of Energy grants DE-FG02-94ER40818 (W.B.) and DE-FG02-86ER40281 (W.Z.).


\newpage
\section{Appendix - Annotated Timeline of Major Developments}
\label{Sec:Appendix}
\begin{addmargin}[3em]{2em}
\begin{itemize}
    
\item[\bf 1930-1935]
\hypertarget{Appendix}{Discovery} of the neutron~\cite{Chadwick:1932ma} leads to the birth of nuclear physics~\cite{Wrob2002}.\\
Realization that cosmic rays include protons and nuclei~\cite{Rossi}.\\  
The combination of cosmic rays and cloud chambers, in use since 1912~\cite{Wilson:1912bva},
become a powerful tool for the study of the interaction with matter of strongly interacting particles~\cite{Rossi1964cosmic}.\\ 
The usefulness of this tool is significantly enhance with the invention of the Geiger counter, coincidence circuit and triggered and multiplate cloud chambers~\cite{Rossi:1982uhu}.

\item[\bf 1938-1943]
Improvements in emulsion technology lead to emulsions which are sensitive to high energy proton tracks~\cite{Shapiro1941}. \\
Observations in cloud chambers and emulsions of multiparticle production, including an 8 prong star observed by Marietta Blau~\cite{Blau1937} in an emulsion placed on a mountain top. Conclusion that there must be some very strong interactions~\cite{Lewis:1948zz}.\\ 
Heitler, Heisenberg and Janossy debate if multiparticle production is a consequence of intra-nuclear cascades~\cite{Hamilton:1943dne,Janossy1943,Heisenberg:1955llt}. \\
Heisenberg proposes colliding ``mesotron" fields as explanation of multiparticle production~\cite{Heis39}. There is general interest if the production process is fast or slow.

\item[\bf 1945-1950]
Clear evidence in a cosmic ray- emulsion event that many particles can be produced in a single pp collision~\cite{Brown:1949mj}.\\
Discovery of the pion~\cite{Lattes:1947mx}.

\item[\bf 1950-1955]
Question is raised if complete stopping in high energy pp collisions can occur and, if so, what are the properties of the hot high energy density matter that is produced. \\
Fermi's thermodynamic model~\cite{Fermi:1950frz}. \\
Landau's relativistic hydrodynamic model~\cite{Landau:1953gs,Landau:1955tlt}. \\
Cosmotron built at BNL~\cite{Cosmotron}. \\
Bevatron built at LBL~\cite{Bevatron}.\\

\item[\bf 1955-1960]
Disovery of the antiproton~\cite{Chamberlain:1955ns}.\\
The HILAC, an 8.5 Mev/u nuclear accelerator is built at LBL to study properties of nuclear matter~\cite{Grunder1973heavy}.\\
Key questions on the mechanism of multiparticle production and the formation time for the creation of new particles are posed~\cite{feuinberg1960multiple}.

\item[\bf 1960-1965]
Quark model proposed to explain flavor structure of 
mesons and baryons~\cite{Zweig:1964jf,Zweig:1964ruk,Gell-Mann:1964ewy}.

\item[\bf 1965-1971]
Discovery of neutron stars~\cite{nStar1967,Hewish:1968bj,lovelace2012discovery}. \\
Resultant interest in dense nuclear matter~\cite{Chapline:1976gy,Baym:1975mf}.\\ 
Observation of large number of resonances leads Hagedorn to conclude that there exists a limiting temperature~\cite{Hagedorn:1967tlw}.\\
Extensive studies of p+A collisions with emulsion detectors~\cite{Fredriksson:1984nb}.\\
There is interest in the formation time of hadrons.\\
Studies of p+A collisions, with $A$ a known element, carried out at the Echo lake cosmic ray facility. 

\item[\bf 1970-1975]
Bevalac begins operations with relativistic nuclear beams~\cite{osti_937059}.\\
p+A studies with up to 200~GeV beams at Fermilab. For the first time the energy of the beam and particle type, 
as well as the target nucleus are all known~\cite{Busza:1975te,Nikolaev:1981dh,Otterlund:1984mao}.\\
Various parton and hydrodynamic models developed to explain the particle production process~\cite{Busza:1975te}. \\
Gottfried introduces the concept of an ``energy flux"~\cite{Gottfried:1974yp} to explain p+A data. \\
There is interest in a baryon-quark phase, dense nuclear matter  and "quark soup"\cite{Itoh:1970uw,Chapline:1976gy,Baym:1976yu,Freedman:1977gz,Cabibbo:1975ig,Collins:1974ky}.\\
The ISR collider begins operations with p+p collisions~\cite{Giacomelli:1979nu,Jacob:1984dy}.\\
Glauber Method developed to determine the impact parameter (or centrality) of  pA and AA collisions~\cite{Bialas:1977pd}.

\item[\bf 1973-1974]
 Quarks, gluons, asymptotic freedom and birth of QCD are game changers~\cite{Politzer:1973fx,Gross:1973id,Politzer:1974fr,Gross:1974cs}! 

\item[\bf 1973-1978]
Participant scaling~\cite{Elias:1979cp} and Cronin effect~\cite{Cronin:1974zm} discovered at Fermilab. \\
EMC effect discovered at SPS~\cite{EuropeanMuon:1983wih}.\\
Interest in quark matter~\cite{Chapline:1976gq},
Lee-Wick Matter~\cite{Lee:1974ma} and other high density exotic nuclear phenomena.\\
At Bear Mountain meeting T.D.Lee argues for the study of heavy ion collisions at ultra high energies~\cite{BearMountain}.\\ 
MIT bag model introduced~\cite{Chodos:1974je}.\\ 
 Suggestion appropriate descriptor is ``quark-gluon plasma"~\cite{Shuryak:1977ut}.

\item[\bf 1975-1982]
Bevalac program pursues studies of compression of nuclear matter and exotic states of nuclear matter~\cite{Stock:2004iim}.\\
Interest in and estimates of the baryon density and energy density that should be achievable in high energy AA collisions~\cite{Goldhaber:1978qp,Busza:1983rj}.\\
ISR program is extended to include $p\mathrm{+}\alpha$ and $\alpha\mathrm{+}\alpha$ collisions~\cite{Jacob:1981zb}.\\
Bjorken preprint on hydrodynamics and energy density in A+A collisions~\cite{Bjorken:1982qr}.\\
Serpukhov studies of $\pi$-A collisions. \\
Influential workshops at GSI~\cite{Bock:1981iyr} and LBL~\cite{Proceedings:1979bna}.

\item[\bf 1980-1983]
Increased interest in quark matter formation and heavy ion collisions~\cite{JACOB1982321}.\\
Bjorken  describes hydrodynamics and energy density in A+A collisions~\cite{Bjorken:1982qr}.\\
Hakone seminar~\cite{Nakai:1980gx,Nakai:1980gw} 
and Bielefeld workshop~\cite{Ericson:1982ji}.\\
CBA terminated~\cite{Samios:2007zz}.\\
NSAC town meeting and go-ahead given  for RHIC construction~\cite{Ludlam:1984klg}.
\end{itemize}
\end{addmargin}



\bibliographystyle{atlasnote}
\bibliography{QGPHistoryBuszaZajc}

\clearpage

\end{document}